\documentclass[sigconf]{acmart}

\usepackage{booktabs}
\usepackage{multirow}
\usepackage{graphicx}
\usepackage{amsmath}
\usepackage{amsfonts}
\usepackage{xcolor}
\usepackage{subcaption}
\usepackage{pgfplots}
\usepgfplotslibrary{groupplots}
\pgfplotsset{compat=1.18}

\definecolor{cdense}{HTML}{4477AA}
\definecolor{cmem}{HTML}{EE6677}
\definecolor{cseq}{HTML}{228833}
\definecolor{cco}{HTML}{AA3377}

\newcommand{\mosaic}{\textsc{Mosaic}}

\newcommand{\todo}[2][]{\textcolor{red}{\textbf{[TODO\if\relax#1\relax\else~(#1)\fi: #2]}}}

\copyrightyear{2026}
\acmYear{2026}
\setcopyright{cc}
\setcctype{by}
\acmConference[RecSys '26]{20th ACM Conference on Recommender Systems}{September 27-October 02, 2026}{Minneapolis, MN, USA}
\acmBooktitle{20th ACM Conference on Recommender Systems (RecSys '26), September 27-October 02, 2026, Minneapolis, MN, USA}
\acmDOI{10.1145/3773078.3831897}
\acmISBN{979-8-4007-2284-4/2026/09}

\begin{document}

\title{\mosaic{}: A Fleet of User Embedding Specialists for Recommendation at Meta}

\author{John Zhiyuan Zheng}
\orcid{0000-0002-7828-0255}
\affiliation{%
  \institution{Meta}
  \city{Menlo Park}
  \country{USA}
}
\email{zhiyuanbj@gmail.com}

\author{Xian Sun}
\orcid{0000-0003-0613-4184}
\affiliation{%
  \institution{Meta}
  \city{Menlo Park}
  \country{USA}
}
\email{xiansun@meta.com}

\author{Xiangyang Mou}
\orcid{0000-0003-3362-8008}
\affiliation{%
  \institution{Meta}
  \city{Menlo Park}
  \country{USA}
}
\email{moux@meta.com}

\author{Yujunrong Ma}
\orcid{0000-0001-5157-837X}
\affiliation{%
  \institution{Meta}
  \city{Menlo Park}
  \country{USA}
}
\email{myjr@meta.com}

\author{Christina You}
\orcid{0009-0000-8573-5564}
\affiliation{%
  \institution{Meta}
  \city{Menlo Park}
  \country{USA}
}
\email{christinaqyou@meta.com}

\author{Michael Jiayuan He}
\orcid{0000-0002-8566-7095}
\affiliation{%
  \institution{Meta}
  \city{Menlo Park}
  \country{USA}
}
\email{michaelhe.official@gmail.com}

\author{Hrishikesh Paranjape}
\orcid{0009-0008-3423-9622}
\affiliation{%
  \institution{Meta}
  \city{Menlo Park}
  \country{USA}
}
\email{hparanjape@meta.com}

\author{Aakarsha Agarwal}
\orcid{0009-0009-0013-4483}
\affiliation{%
  \institution{Meta}
  \city{Menlo Park}
  \country{USA}
}
\email{aakarsha@meta.com}

\author{Hong Li}
\orcid{0009-0008-8299-2328}
\affiliation{%
  \institution{Meta}
  \city{Menlo Park}
  \country{USA}
}
\email{hongli@meta.com}

\renewcommand{\shortauthors}{Zhiyuan Zheng et al.}

\begin{abstract}

User representation is one of the highest-leverage modeling problems in industrial recommendation systems: a single advancement in how users are encoded can propagate across retrieval, ranking, and integrity tasks at platform scale. Prior industrial user representation work builds either a single user model that emits one or more embedding vectors or a shared backbone with task-specific adaptation. In this paper, we present \mosaic{}, a foundational user modeling platform that employs a fleet of specialists to learn user embeddings. The fleet comprises four architecturally diverse model families -- \emph{memorization-driven}, \emph{dense-heavy}, \emph{sequential-based}, and \emph{CoTrain models} -- each focusing on a distinct facet of user behavior. We developed MRM (Multi-task Relations Mining) and CRL (Cosine
Redundancy Loss) techniques to maximize the marginal information contribution of each new specialist. We also introduce \emph{CoEval} and \emph{User Tower Zero-Out}, new logging-free embedding evaluation framework that improves development velocity while preserving downstream-aligned accuracy. Our hybrid CPU/GPU, online-and-offline serving stack allows each specialist to choose the adequate serving strategy to meet the freshness, latency, and computational requirements. \mosaic{} delivers consistent and significant offline NE improvements in addition to online gains.

% Most prior industrial user-representation work builds either a single user model, a single model that emits multiple user vectors, or a shared foundation/backbone with task- or surface-specific adaptation. In contrast, Mosaic treats heterogeneous user embeddings as a managed fleet of specialist models under a common downstream contract.

\end{abstract}




\maketitle

% =============================================================================
% Section 1: Introduction
% =============================================================================

\section{Introduction}
\label{sec:intro}

% User representation is a high-leverage problem in modern recommendation systems: user-level signals---including sparse IDs, categorical features, behavior sequences, and learned embeddings---capture distinct facets of behavior but vary widely in coverage. This unevenness is especially visible for cold-start users, whose limited histories often lead to worse model performance. Learning from broad, heterogeneous user signals is therefore critical: a more holistic user representation can improve downstream retrieval, prediction, and integrity tasks.

User representation is a central problem in modern recommendation systems. User-level signals, including sparse IDs, categorical features, engagement sequences, and learned embeddings, capture different aspects of user preference, but their coverage is uneven across cohorts. This is especially visible for cold-start users, whose limited histories often lead to worse model performance. Improving how these heterogeneous signals are combined into a holistic user representation can therefore benefit downstream retrieval, prediction, and integrity tasks.

% A natural approach is to train a dedicated user embedding model that focuses exclusively on user features, distilling high-dimensional, heterogeneous user data into compact embedding vectors consumed as input features by downstream models. Compared with learning user representations within each downstream ranker directly, a dedicated embedding model offers several advantages: it provides a free model scale-up without inflating downstream resource usage; its decoupled training architecture can leverage data and supervision signals that downstream tasks cannot easily incorporate; and the resultant embeddings can be shared across many downstream models at no additional inference capacity.

A common approach is to train a dedicated user embedding model that compresses heterogeneous user features into compact vectors for downstream rankers. Compared with learning user representations directly inside each ranker, this decoupled design has several advantages. It provides a free model scale-up without inflating downstream computational and memory usage, can leverage data and supervision signals that downstream tasks cannot easily incorporate, and makes the resulting embeddings reusable across many downstream consumers.

% A growing body of industry work has explored this direction, but existing approaches remain limited in complementary ways. Early user embedding systems such as YouTube DNN~\cite{covington2016deep} and PinnerSage~\cite{pal2020pinnersage} learn single-surface, single-architecture representations that compress all user knowledge into one model. More recent foundation model efforts---PinFM~\cite{chen2025pinfm}, 360Brew~\cite{firooz2025_360brew}, and Tencent's Large User Model~\cite{yan2025lum}---scale to larger architectures and cross-surface pre-training, but adopt a monolithic design that forces a single model to balance various signal types and learning objectives. HSTU~\cite{zhai2024actions} embeds user representation as an internal module of the ranking model, leaving the learned representation entangled with the ranker and therefore unable to be decoupled and shared with other downstream consumers. Multi-task architectures like PLE~\cite{tang2020ple} and HoME~\cite{wang2024home} introduce task-specific expert routing but still operate within a single model architecture, and do not explore whether different types of user knowledge would benefit from architecturally distinct specialists. What unites these prior approaches is a common design choice: a \emph{single} user model, whether monolithic or task-shared on a common backbone. We explore an orthogonal direction---a managed \emph{fleet of architecturally specialized embedding models}---and discuss its tradeoffs against the single-model paradigm in Section~\ref{sec:design_principles}.

A growing body of industry work has pursued this direction. Early systems learn user representations from interaction history through deep retrieval models~\cite{covington2016deep, yi2019sampling}, while Pinterest's work adds clustering-based multi-vector representations and batch-refreshed sequence embeddings~\cite{pal2020pinnersage, pancha2022pinnerformer}. Multi-interest methods produce multiple vectors per user through routing, interest extraction, or condition-specific modules~\cite{li2019multi, cen2020controllable, fan2025pinterest_multiembed}. Recent foundation-scale designs further scale user-side modeling with large pretrained sequence models, generative paradigms, per-application fine-tuning, or shared backbones with lightweight experts~\cite{chen2025pinfm, yan2025lum, li2025foundation_expert}. Despite their differences, these approaches largely center user representation around a single user-side architecture or a shared backbone, even when producing multiple vectors per user. We instead explore a different design decision: a heterogeneous library of specialist embedding models, and discuss the tradeoffs against the single-backbone paradigm in Section~\ref{sec:design_principles}.

% We instantiate this design in \mosaic{}, a foundational user-modeling platform. The fleet comprises four specialist categories, each architecturally matched to a distinct facet of user behavior: (1)~\textit{memorization driven} models that leverage large-scale hash tables to capture fine-grained per-user preferences; (2)~\textit{dense-heavy} models employing architectures such as Multi-Head Talking Attention (MHTA), Squeeze-and-Excitation Dot (SEDot), and Wukong for adaptive, nonlinear feature interactions over aggregated user counters; (3)~\textit{sequential} models built on HSTU~\cite{zhai2024actions} with Mixture-of-Experts (MoE) routing for capturing temporal interest drift, bursty engagement, and social influence patterns; and (4)~\textit{Co-Train} models with end-to-end gradient flow from downstream rankers for task-aligned representations.

We instantiate this design in \mosaic{}, a user-modeling platform built around four specialist families, each matched to a distinct facet of user behavior:  (1)~\textit{Memorization-driven} models use large hash tables to capture fine-grained per-user preferences. (2)~\textit{Dense-heavy} models learn nonlinear interactions over aggregated user counters, with architectures such as MHTA~\cite{shazeer2020talkingheads}, SEDot~\cite{hu2018senet}, and Wukong~\cite{wukong}. (3)~\textit{Sequential} models build on HSTU~\cite{zhai2024actions} with MoE routing to capture temporal dynamics in user behavior. (4)~\textit{CoTrain} models receive end-to-end gradients from downstream rankers to learn task-aligned representations.

To serve these specialists at scale, we develop a hybrid serving stack that supports both CPU and GPU online inference, maximizing the utilization of spare computational resources. In particular, we optimize GPU inference to reduce online serving capacity while meeting latency targets. We developed both online and offline inference of user embeddings, which offers flexibility regarding feature freshness and efficiency. This trade-off extends the applicability of Mosaic to use cases with limited computational resources. Based on our experiments, \mosaic{} delivers consistent offline Normalized Entropy (NE) gains and statistically significant online wins.

Our key contributions are:
\begin{itemize}
    \item \textbf{Fleet-of-specialists design.} We present the design rationale and empirical validation for building a diverse fleet of user embedding models with heterogeneous architectures, each matched to a distinct inductive bias---a paradigm not systematically explored in prior work (\S\ref{sec:system:fleet}).
    \item \textbf{Cross-surface data for holistic user representation.} We leverage data from multiple surfaces for both model supervision and input features, enabling each specialist to learn cross-surface user behavior that single-surface models cannot capture (\S\ref{sec:system:fleet}).
    \item \textbf{Maximizing unique user knowledge distillation.} We introduce Multi-task Relations Mining (MRM) and Cosine Redundancy Loss (CRL) techniques to maximize the marginal information contribution of each new specialist, addressing a fundamental challenge in multi-embedding systems (\S\ref{sec:challenges:overlap}).
    \item \textbf{Hybrid serving for resource-efficient deployment.} We develop a hybrid serving stack that combines online and offline, CPU and GPU paths, balancing freshness, latency, and compute usage on a per-specialist basis (\S\ref{sec:system:serving}).
    \item \textbf{Logging-free embedding evaluation.} We introduce \emph{CoEval} and \emph{User Tower Zero-Out}, two evaluation methods that measure an embedding's downstream-aligned contribution directly from a trained checkpoint--bypassing the standard feature-logging pipeline and improving iteration velocity by 3--5$\times$ while preserving accuracy (\S\ref{sec:challenges:eval}).
\end{itemize}

% The remainder of the paper is organized as follows. Section~\ref{sec:related} positions \mosaic{} against prior work in user representation and foundation models for recommendation. Section~\ref{sec:system} describes the system, covering the experimental setting, design principles, the four specialist architectures, and the hybrid serving stack. Section~\ref{sec:challenges} presents the techniques we developed to maximize unique user knowledge across specialists and to evaluate embedding quality without the feature-logging pipeline. Section~\ref{sec:deployment} reports offline NE improvements and online A/B results, alongside specialist-family contribution and sequential-scaling ablations. Section~\ref{sec:conclusion} concludes with the broader applicability of the design and key takeaways.

The remainder of the paper is organized as follows. Section~\ref{sec:related} reviews prior work in user representation for recommendation. Section~\ref{sec:system} presents the system design, specialist architectures, and serving stack. Section~\ref{sec:challenges} describes user knowledge overlap reduction and logging-free evaluation. Section~\ref{sec:deployment} reports offline and online A/B results, specialist-level contributions, and sequential-scaling ablations. Section~\ref{sec:conclusion} concludes with the broader applicability of the design and key takeaways.

% =============================================================================
% Section 2: Related Work
% Four thematic paragraphs framed around operational specialization.
% Logic-driven: describe approaches, not model names.
% =============================================================================

\section{Related Work}
\label{sec:related}

\textbf{Dedicated User Embedding Models.}
Industrial systems have long used dedicated user-side models whose outputs can be reused by downstream retrieval or ranking systems. Early works relied on using two-tower model architectures to learn user representations for retrieval~\cite{covington2016deep, yi2019sampling}. Later systems expanded this direction with behavioral clustering~\cite{pal2020pinnersage}, batch-refreshed sequence encoders~\cite{pancha2022pinnerformer}, multi-vector interest representations~\cite{li2019multi, cen2020controllable, fan2025pinterest_multiembed}, and foundation-scale user-side models~\cite{chen2025pinfm, yan2025lum}. More recent foundation-plus-expert designs combine a shared cross-surface backbone with surface-specific experts~\cite{li2025foundation_expert}. These systems establish the value of user-side representations, but they are typically organized around a particular model family or shared backbone, thus they cannot fully distill the cross-surface user behavior. \mosaic{} takes a different route: it manages user representation as a fleet of specialist embedding models: each specialist is focusing on learning a distinct facet of user behavior through an independently trained model with its own architecture, training data, refresh cadence, and serving regime.

\textbf{In-Model User Representations.}
User representations can also be learned within retrieval, ranking, or end-to-end recommendation models. This includes classical latent-factor models~\cite{koren2009matrix}, deep rankers with sparse/dense feature interactions~\cite{cheng2016wide, wang2017dcn, wukong, zhu2025rankmixer}, target-attention and long-history models that compute candidate-conditioned user interests~\cite{zhou2018deep, dien, pi2020search, pi2023twin, si2024twinv2, chai2025longer, guan2025makelong}, and sequential recommenders trained with next-item, masked-item, self-supervised, or contrastive objectives~\cite{sasrec, sun2019bert4rec, zhou2020s3rec, xie2022cl4srec}. Multi-task and multi-domain rankers further specialize representations through gated experts, shared/task-specific modules, or personalized routing~\cite{mmoe, tang2020ple, star, wang2024home, chang2023pepnet}. Recent generative recommenders couple representation learning to retrieval or ranking through sequential transduction, unified generation, or semantic-ID decoding~\cite{zhai2024actions, deng2025onerec, zhou2025onerec_v2, rajput2024tiger}, while hybrid systems combine in-ranker sequence modules with batch-generated user embeddings~\cite{xia2023transact, xia2025transactv2}. In these approaches, user representation is usually coupled to a specific objective, feature space, and serving graph, which limits user-side scaling, cross-surface signal transfer, and reuse across downstream consumers. \mosaic{} avoids these constraints through a fleet of dedicated specialist embedding models that are trained and served independently, then exposed as reusable features to multiple downstream consumers.

% \textbf{Sequential and long-history modeling.}
% Sequence modeling is a trending topic in user representation. Target-attention methods extract per-target interests from behavior histories~\cite{zhou2018deep, dien}, while retrieve-then-attend and long-context architectures scale this idea to longer histories~\cite{pi2020search, pi2023twin, si2024twinv2, chai2025longer, guan2025makelong}. Self-attentive models learn user representations through next-item or masked-item objectives~\cite{sasrec, sun2019bert4rec}, and generative recommenders reformulate ranking as sequential transduction~\cite{zhai2024actions}. \mosaic{} uses this line of work as one specialist to enrich the model family.

\textbf{Serving and evaluation infrastructure.}
Large-scale user embeddings also impose stringent infrastructure requirements, spanning storage, freshness, serving cost, and evaluation. Prior systems have addressed parts of this problem by real-time embedding tables and hybrid sparse–dense training at scale~\cite{monolith, lian2022persia}. \mosaic{} builds on these primitives but faces a different operational problem: serving a heterogeneous fleet of models that vary in the needs of freshness, latency, cost, and inference hardware. Its serving stack supports different resource landscapes via CPU online inference, GPU online inference, and GPU offline inference. Our logging-free evaluation framework, the CoEval, simulates the online serving environment locally and allows fast iterations on the fleet of embeddings.

\begin{figure*}[!ht]
    \centering
    \includegraphics[width=0.75\textwidth]{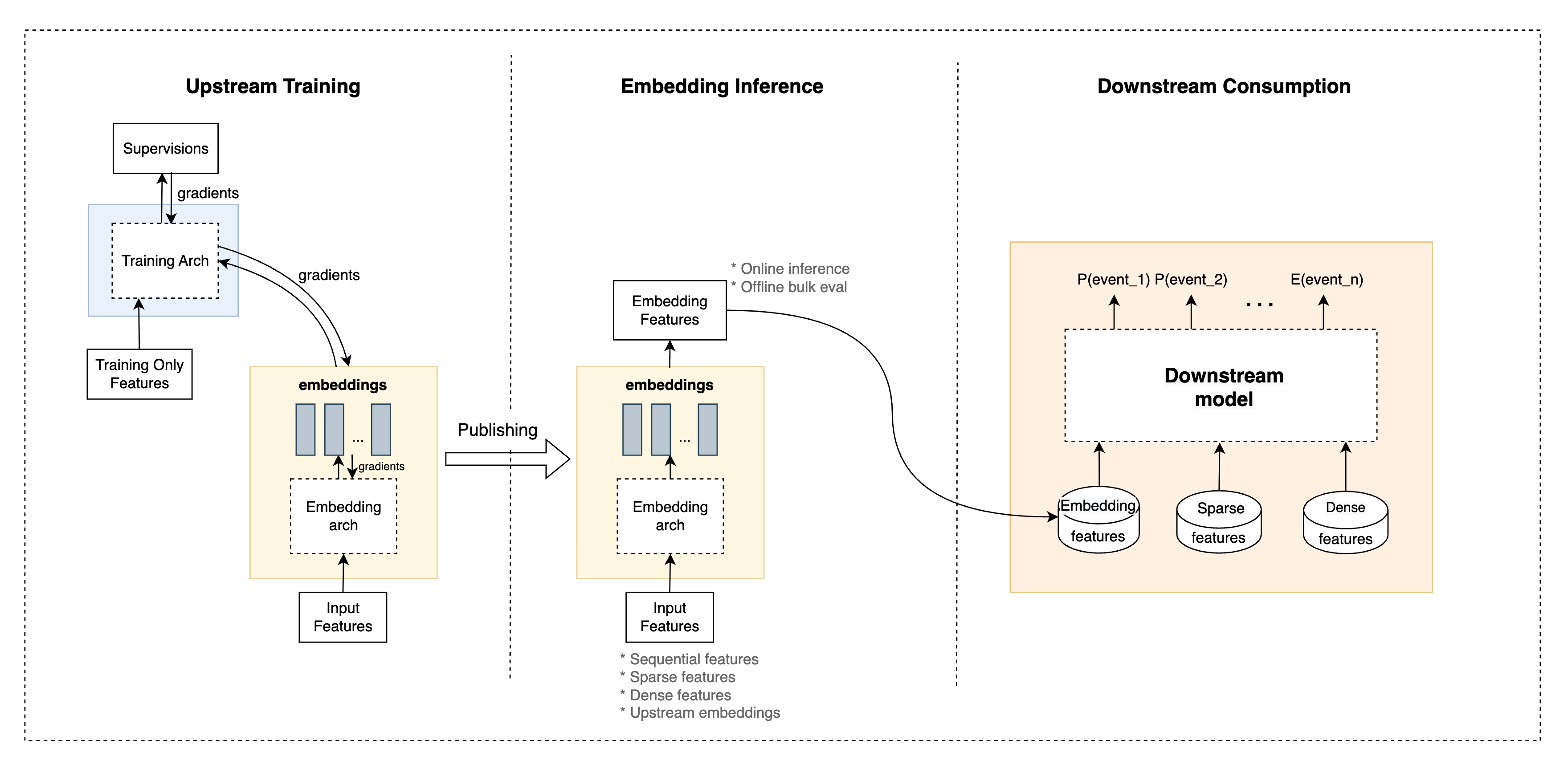}
    \caption{High-level \mosaic{} system overview.}
    \Description{System overview diagram showing the Mosaic fleet architecture: training pipelines on the left feed embedding generation, which in turn populates two parallel serving paths (an online GPU-based path and an offline batch path); both serving paths expose user embeddings under a uniform contract to downstream ranking and retrieval consumers shown on the right.}
    \label{fig:architecture}
\end{figure*}

\section{The Mosaic System}
\label{sec:system}

\subsection{System Overview}
\label{sec:overview}

The \mosaic{} system operates in three stages, as illustrated in Figure~\ref{fig:architecture}: upstream model training, embedding inference, and downstream consumption. During upstream training, each specialist model is trained with a full training graph consisting of two parts: (1) a training-only architecture ingesting non-user features and supervised by engagement tasks, and (2) a user tower, which ingests heterogeneous user features including sparse IDs, dense counters, and pre-trained embeddings. The training-only architecture provides multi-task gradient signals back to the user tower, and the user tower learns to compress these signals into one or more user embedding vectors. 

At inference time, only the user tower is deployed. Given user-side input features, it generates embeddings either online for freshness sensitive use cases or through offline bulk inference for embeddings with low freshness requirements. Because the training-only architecture is not needed after the training stage, the serving footprint is substantially smaller than the full training graph. New specialists enter the inference path through model configuration, binary loading, and staged traffic ramping before downstream consumers receive the new embedding.

Downstream rankers consume \mosaic{} outputs as standard embedding features, alongside existing user, item, and context features. This design lets a ranker benefit from a larger upstream user-modeling system without enlarging its own architecture or serving graph. Both the training-only architecture and the user tower are agnostic to the downstream ranker's architecture: SNNs~\cite{naumov2019deep}, Transformers, and dense feature-interaction networks are all supported in the \mosaic{} framework. Therefore, depending on its role in the fleet, each \mosaic{} specialist has a distinct user tower and training-only design that differs in model architecture, supervisory tasks, and/or feature sets. This configurable setup underpins the specialist designs described in Section~\ref{sec:system:fleet}.

We evaluate \mosaic{} embeddings with downstream ranking metrics, including NE, Feature Importance (FI), and online A/B metrics. Since upstream NE can diverge from downstream impact, we use the downstream ranker's NE as the primary offline measure of embedding quality. The standard evaluation path is slow, requiring deployment, feature logging, and downstream retraining before impact can be measured; this motivates the logging-free CoEval framework in Section~\ref{sec:challenges:eval}.

%% ---------------------------------------------------------------------------
%% 3.1 Problem Setting and Design Principles
%% ---------------------------------------------------------------------------

\subsection{Design Principles}
\label{sec:design_principles}
\paragraph{Dedicated user embedding model vs.\ scaling downstream user features.}
A dedicated upstream user model is qualitatively different from simply exposing more user-side features to each downstream ranker. It enables three capabilities that downstream feature scaling cannot provide on its own. First, it amortizes user-side modeling capacity: downstream rankers receive compact learned embeddings, while the heavier training graph remains outside their serving path. Second, the upstream model can use cross-surface features, auxiliary labels, and supervision tasks that are difficult to add directly to every downstream ranker. Third, the same learned embeddings can be reused by multiple consumers, avoiding repeated feature engineering and model-capacity increases across teams. Empirically, we also observe that a dedicated user tower can add signal value even when it overlaps with features already available to the downstream model, suggesting that the upstream model learns complementary user representations rather than simply duplicating downstream computation.

\paragraph{Fleet of specialists vs.\ a single generalist.}
Given a dedicated upstream user-modeling layer, the next design choice is whether to build one general-purpose user model or multiple specialists. We choose a specialist fleet because the dominant user-signal families in our setting have different statistical structure and serving requirements. Sparse and categorical features favor memorization-heavy models; dense behavioral counters benefit from nonlinear feature interaction; sequential actions require temporal modeling; and downstream-specific objectives benefit from task-aligned training. Table~\ref{tab:taxonomy} summarizes this mapping between signal types and inductive biases. A single generalist model can represent these signals, but it must share capacity and optimization across heterogeneous objectives. In contrast, a fleet lets each specialist optimize on one signal dimension. In a prototype comparison with matched data and compute, a fleet of smaller specialists produced stronger embedding quality than a large generalist model. Table~\ref{tab:fleet_vs_mono} summarizes the resulting tradeoffs in learning focus, development velocity, adoption cost, and failure isolation.

\begin{table}[t]
\caption{Mosaic Fleet: model categories emerge from user signal types and optimal inductive bias}
\label{tab:taxonomy}
\footnotesize
\setlength{\tabcolsep}{3pt}
\begin{tabular}{@{}p{2.0cm}p{1.3cm}p{1.9cm}p{2.4cm}@{}}
\toprule
\textbf{Model Category} & \textbf{Feature Type} & \textbf{Inductive Bias} & \textbf{Example User Signal} \\
\midrule
Memorization-driven & Sparse list & Per-user lookup; memorize preference & Liked sports, shared food post \\
\midrule[0.02em]
Dense-heavy & Dense counters & Nonlinear counter combinations & 30d avg 12 likes/day, 80\% mobile \\
\midrule[0.02em]
Sequential &  History seq. & Attention over temporal deps. & MMA, then Wrestling, then Jiu-jitsu \\
\midrule[0.02em]
CoTrain & Cross-surface & E2E gradient from downstream task & Feed emb.\ learned from Notif pClick \\
\bottomrule
\end{tabular}
\end{table}

% \begin{table}[t]
% \caption{Monolithic FM vs.\ Fleet of Specialists: tradeoffs across key axes.}
% \label{tab:fleet_vs_mono}
% \footnotesize
% \setlength{\tabcolsep}{3pt}
% \begin{tabular}{@{}p{1.8cm}p{2.6cm}p{2.7cm}@{}}
% \toprule
% \textbf{Axis} & \textbf{Single Monolithic FM} & \textbf{Fleet of Specialists} \\
% \midrule
% Learning focus & One model balances memorization, sequential, and dense generalization in shared capacity & Each specialist owns one facet---no capacity competition \\
% \midrule[0.02em]
% Dev velocity & Slow---one model blocks all & Fast---parallel dev across specialists \\
% \midrule[0.02em]
% Adoption cost & High---KD/TL pipeline per consumer & Low---plug-and-play embedding feature \\
% \midrule[0.02em]
% Failure blast radius & Wide---single point of failure & Small---isolated per specialist \\
% \bottomrule
% \end{tabular}
% \end{table}

\begin{table}[t]
\caption{Qualitative tradeoffs between a general-purpose user model and a specialist fleet.}
\label{tab:fleet_vs_mono}
\footnotesize
\setlength{\tabcolsep}{3pt}
\begin{tabular}{@{}p{2cm}p{3cm}p{3cm}@{}}
\toprule
\textbf{Design Dimension} & \textbf{A single generalist} & \textbf{The fleet of Specialists} \\
\midrule
Signal focus &
Shared capacity must cover sparse, dense, and sequential signals &
Each specialist focuses on a narrower signal family and inductive bias \\
\midrule[0.02em]
Development velocity &
Changes are coupled through one model release cycle &
Specialists can be developed and validated independently \\
\midrule[0.02em]
Failure isolation &
Quality or serving regressions can affect the shared model broadly &
Regressions can be isolated to the affected specialist \\
\bottomrule
\end{tabular}
\end{table}

%% ---------------------------------------------------------------------------
%% 3.2 The Fleet
%% ---------------------------------------------------------------------------

\subsection{Specialist Architectures}
\label{sec:system:fleet}

\begin{figure*}[!ht]
    \centering
    \includegraphics[width=\textwidth]{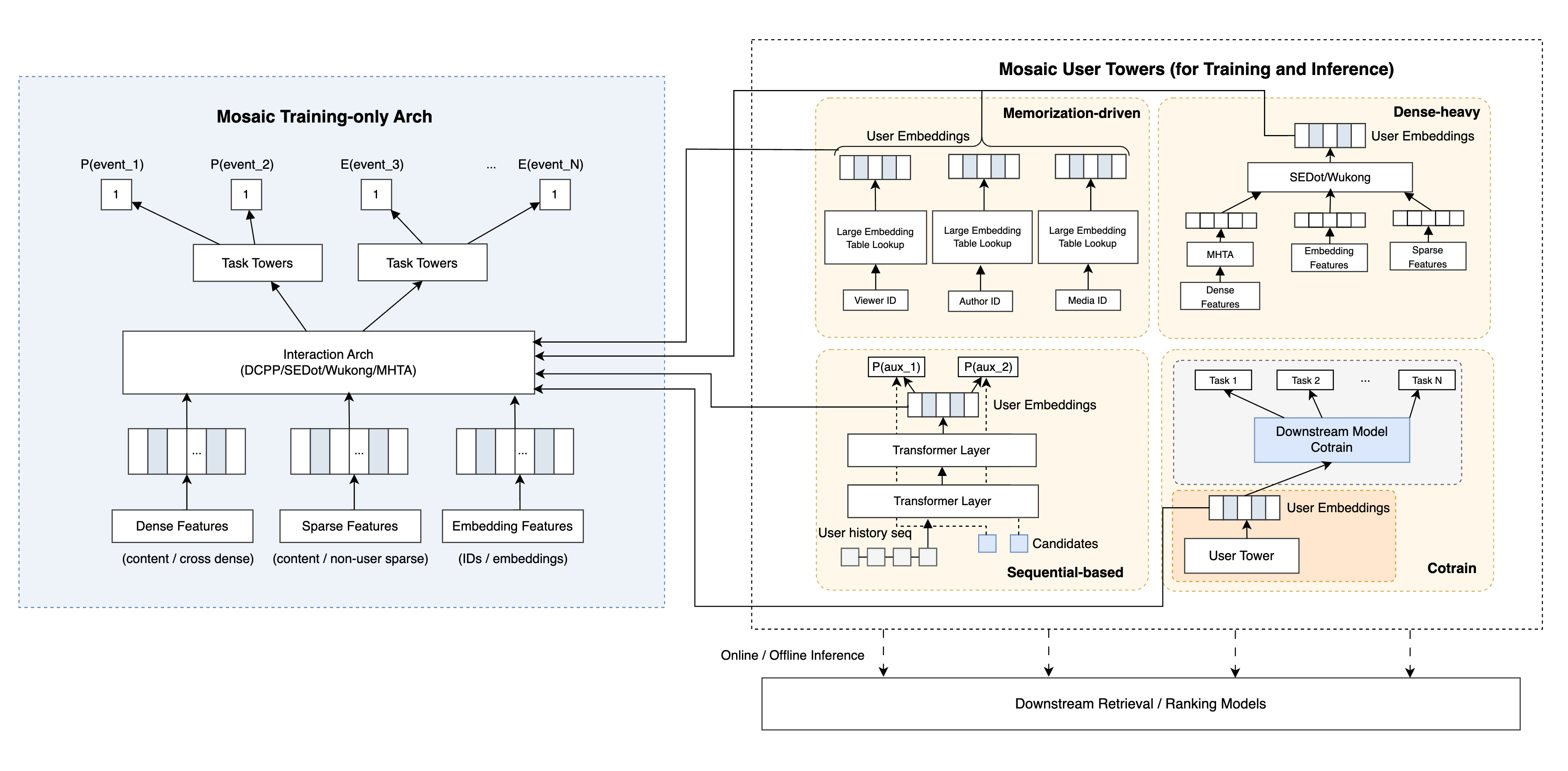}
    \caption{Model architecture of the \mosaic{} fleet, comprising a training-only architecture (left) and four user-tower specialist families (right). For simplicity, we draw all embeddings pointing to one training-only block. In practice, each \mosaic{} specialist has its own distinct training-only architecture.}
    \Description{Architecture diagram with two main blocks. The left block, "Main Recommendation Architecture (Shared Training Backbone)", shows player inputs (P_1 through P_N) feeding an attention layer ATT, which feeds an Interaction Arch labeled "Shared Representation / Interaction Layers", which in turn ingests dense features, sparse features, and embedding features. The right block, "Model Fleet", is divided into four sub-blocks: CoTrain Models with a shared backbone and Task 1 / Task 2 / Task N heads driving multiple downstream training objectives; Sequential-based Models showing player history fed through transformer layers for long-sequence interest history modeling; Dense-heavy Model showing a user tower, item tower, and interaction layer leveraging rich dense features for ranking; and Memorization-driven Models showing lookup of voter, friend, and watchlist signals to recall learned user preference via memory lookup. Both blocks feed a Score Fusion / Calibration / Ranking stage that produces the Final Recommendation Output.}
    \label{fig:model_arch}
\end{figure*}

The fleet comprises four specialist families, each matched to a distinct objective: \emph{memorization-driven} (sparse features), \emph{dense-heavy} (aggregated counters), \emph{sequential} (action histories), and \emph{CoTrain} (downstream engagement). Figure~\ref{fig:model_arch} illustrates the design of these four specialist architectures.

\subsubsection{Memorization-Driven Specialist}
\label{sec:system:memorization}
Our memorization specialists rely on large-scale embedding hash tables to capture behavioral patterns. Factoring large tables into an upstream specialist condenses the memorized signal into a compact embedding before it reaches the downstream model, allowing the table's computational cost to be amortized across multiple consumers. However, supporting these tables at scale (i.e., hundreds of gigabytes of states) without impacting production QPS requires coordinated optimizations across both training and serving.

We support extra-large embedding tables with three techniques across training and serving. \emph{Table-Batched Embedding (TBE)} fuses many lookup operations into one kernel, tiers rows across GPU HBM, CPU DRAM, and SSD by access frequency, and quantizes per table down to INT4. For training, \emph{TorchRec row-wise sharding} splits each table across GPU HBM so every lookup hits one shard. For serving, \emph{distributed inference} uses a heterogeneous plan that keeps hot tables in GPU HBM and shards cold tables across CPU DRAM on multiple machines. Since most of the per-user signal is already captured by the hash tables, we keep the specialist's interaction network deliberately simple, consuming only a small set of top FI dense features together with the hash-table embeddings.

\subsubsection{Dense-Heavy Specialist}
\label{sec:system:dense}
Dense-heavy specialists ingest rich aggregated counter features that span multiple time windows (e.g., 1d, 7d, 28d), engagement types (e.g., likes, comments, shares), surfaces, and action granularities, covering user-level, user--content cross, and content-level signals. Because dense counters benefit more from non-linear feature interaction than from memorization, we design the specialist to learn rich feature crosses and emit multiple embeddings for downstream consumers. We evaluate with multiple model architectures: \emph{MHTA} (Multi-Head Talking Attention~\cite{shazeer2020talkingheads}), which applies multi-head and talking-head gating over per-head DeepCrossNet~\cite{wang2017dcn} blocks; \emph{SEDot} (Squeeze-and-Excitation Dot Networks), which combines sparse-feature dot-product attention with SE-style~\cite{hu2018senet} re-weighting on dense features; and \emph{Wukong}~\cite{wukong}, which comprises stacks of learnable Factorization-Machine layers with a clean NE scaling law. Beyond architecture, we find it important to align the upstream supervision heads with the downstream ranker's top tasks to ensure embedding quality.

\subsubsection{Sequential Specialist}
\label{sec:system:sequential}

The sequential specialist models user behavior dynamics from long cross-platform engagement histories, which are merged into a single chronological stream. It builds on HSTU~\cite{zhai2024actions} with three modifications. First, we prepend contextual tokens. Dense embeddings from cross-surface signals are projected to the sequence dimension and added as prefix tokens. Second, we interleave User Interaction History (UIH) tokens. Each user action contributes content embeddings and metadata embeddings, such as action type, watch time, and temporal position. Since content and metadata are aggregated separately and occupy separate token positions, this version of attention 
disentangles content signals from behavior— our models can now attend to both "what items were consumed" and "how they were consumed". Compared to the sequence length of $N$ in standard HSTU, contextual tokens and UIH interleaving produces an effective sequence length of $C + 2N$, where $C$ is the contextual prefix length and $N$ is the history length. Third, we use hierarchical attention. An L1 self-attention stack processes the full interleaved sequence, and a separate L2 cross-attention layer scores target items against the L1 output.

Two practical issues materially affect embedding quality. First, training and serving used different sequence-length regimes: training uses long histories, while serving requires shorter sequences to meet latency budgets. We address this with \emph{stochastic-length training}, which randomly truncates each training input to a length sampled from the serving-time distribution. This exposes the model to the lengths it will see in production rather than only the maximum training length. Second, timestamp inconsistencies in the logging pipeline occasionally placed user actions out of order. Because causal masking and positional attention assume chronological order, we enforce strict ordering during feature preprocessing. We also studied MoE scaling in the HSTU backbone. A learned router assigns each token to the top-$k$ of $E$ experts, increasing parameter capacity through sparse activation without proportional inference cost. We report the NE and QPS tradeoffs from scaling sequence length, expert count, embedding dimension, and related factors in Section~\ref{sec:deployment:offline} and Table~\ref{tab:seq_scaling}.

\subsubsection{CoTrain Specialist}
\label{sec:system:cotrain}
The first three specialists are trained only through separate upstream training architecture and are not tied to a specific downstream consumer. This makes the resulting embeddings reusable across many rankers, but it can also leave them misaligned with a particular surface, especially when that surface has a different data distribution or supervision target. The \emph{CoTrain specialist} addresses this by introducing a second training architecture during upstream training: a copy of the target downstream ranker, trained on its own data, alongside the standard \mosaic{} training-only architecture. Both architectures share the same user tower, so the embedding is shaped by \mosaic{}'s supervisory objectives along with the target surface objectives.

The CoTrain specialist therefore brings downstream supervision into upstream training. Gradients from the downstream-ranker copy flow end-to-end into the user tower, aligning the embedding with its target consumer before deployment. Serving remains decoupled: at inference time, the downstream rankers consume the embedding through the same method as the other specialists. The framework supports two modes: CoTrain and CoEval. In \emph{trainable} mode, CoTrain updates the user tower with gradients from both training architectures and ships the resulting co-aligned embedding. In \emph{frozen} mode, CoEval keeps \mosaic{}'s training-only arch and user tower fixed and trains only the downstream copy, which we use as an evaluation accelerator (Section~\ref{sec:challenges:eval}). In practice, co-training with a target surface consistently improves embedding quality on that surface, even when the upstream features and supervision are unchanged.
\subsection{Hybrid Serving}
\label{sec:system:serving}

The fleet is served through a hybrid strategy that combines three paths: \emph{CPU online inference}, \emph{GPU online inference}, and \emph{GPU offline inference}. All three paths produce embeddings in the same format, so downstream rankers consume the results in the same way regardless of how the embedding is generated. The hybrid design allows us to balance multiple deployment considerations, including freshness, GPU serving resources, and latency budget. With three paths available, we can pick the configuration that best fits each specialist's decay profile and computation budget.

\textbf{Online inference optimizations.}
GPU online inference is the most resource-intensive path and can bottleneck experimentation velocity, so we minimize its computational usage through a stack of compiler, runtime, and caching layers: 

\begin{itemize}
    \item \textbf{AOTInductor (AOTI) compilation} bypasses the Python runtime through aggressive operator fusion.
    \item \textbf{Predictor tuning} adjusts batch coalescing, thread-pool sizing, and intra/inter-op parallelism.
    \item \textbf{Model splitting} routes \mosaic{}'s heterogeneous sparse-embedding and dense-transformer paths through specialized execution.
    \item \textbf{A look-aside memcache layer} with a 2-hour TTL serves repeat lookups directly from cache, exploiting the slow temporal drift of user embeddings.
\end{itemize}

Table~\ref{tab:gpu_serving} reports cumulative GPU serving reductions as optimizations are added. AOTI with model splitting reduces GPU serving usage by 56\%. Adding memcache on top brings the cumulative reduction to 79\%, while also improving first-story serving latency by 0.14\%.

% \textbf{Online inference optimizations.}
% GPU online inference is the most expensive serving path and often limits experiment throughput. We reduce its cost with compiler, runtime, and caching optimizations: AOTInductor (AOTI) compilation removes Python overhead and fuses operators; predictor tuning adjusts batch coalescing, thread pools, and intra-/inter-op parallelism; model splitting sends \mosaic{}'s sparse-embedding and dense-transformer paths through separate optimized execution paths; and a look-aside memcache layer with a 2-hour TTL serves repeated user lookups from cache, taking advantage of the relatively slow drift of user embeddings. Table~\ref{tab:gpu_serving} reports the cumulative effect of these optimizations, where each row is applied on top of the previous one. The full stack reduces GPU serving capacity by $\sim$79\% and improves first-story serving latency by 0.14\%.

% \begin{table}[h]
% \caption{GPU serving optimization improves both compute cost and serving latency.}
% \label{tab:gpu_serving}
% \small
% \centering
% \begin{tabular}{@{}lcc@{}}
% \toprule
% \textbf{Opt. Methods} & \textbf{GPU Serving Cost} & \textbf{Latency} \\
% \midrule
% AOTI + model split  & $-$56\%  & --        \\
% Memcache            & $-$53\%  & -0.14\%   \\
% \midrule
% Combined            & $-$79\%  & -0.14\%   \\
% \bottomrule
% \end{tabular}
% \end{table}

\begin{table}[h]
\caption{GPU serving optimization improves both compute cost and serving latency.}
\label{tab:gpu_serving}
\small
\centering
\begin{tabular}{@{}lcc@{}}
\toprule
\textbf{Opt. Methods} & \textbf{GPU Serving Usage} & \textbf{Latency} \\
\midrule
Baseline                         & --        & --        \\
AOTI + model split               & $-$56\%   & --        \\
AOTI + model split + Memcache    & $-$79\%   & $-$0.14\% \\
\bottomrule
\end{tabular}
\end{table}

\textbf{Offline inference.}
Large specialists, especially the CoTrain variants, are deployed through an offline inference pipeline: the user tower is extracted from the trained model and applied to the full user population via recurring batch inference, and the resulting embeddings are written to persistent storage and served through the embedding-indexing service.

% =============================================================================
% Section 4: Operating the Fleet
% Two operating challenges only. The User SID quality diagnostic lives in
% Section 5.3 as a complementary qualitative quality signal.
% =============================================================================

\section{Technical Challenges and Solutions}
\label{sec:challenges}

Throughout the development of \mosaic{}, we have encountered a wide range of technical challenges. In this section we discuss the two most central challenges, along with the solutions we developed for them: \emph{maximizing the unique user knowledge} distilled by each new specialist as the fleet grows (Section~\ref{sec:challenges:overlap}), and \emph{optimizing embeddings evaluation} or measuring our signals' downstream impact without going through the feature-logging pipeline (Section~\ref{sec:challenges:eval}).

\subsection{Maximizing Unique User Knowledge}
\label{sec:challenges:overlap}

As we implement successive specialists, we observe \emph{diminishing marginal gains}: each new embedding's lift on downstream metrics shrinks once it is added on top of the existing fleet, even when the same embedding is strong in isolation. The cause is \emph{information redundancy}. Successive specialists trained on overlapping input features and similar supervision signals converge to similar representational subspaces, limiting the novel information each new model contributes. We developed multiple complementary techniques to mitigate this as follows.

\subsubsection{Multi-Task Relation Mining (MRM)}

The downstream ranking model optimizes over multiple engagement prediction tasks, each modeled independently. Mining and exploiting \emph{inter-task relations} yields supervision signals that capture user behavioral patterns not well represented by any single task alone. Intuitively, composite cross-task labels can differentiate fine-grained user intents that individual tasks blur together: a user who both shares and likes a post may signal a different degree of interest than one who only likes, and treating these as the \emph{same} positive label loses that distinction.

Although the upstream model is multi-task, each per-task head supervises only its \emph{marginal} label distribution; the joint distribution across tasks receives no direct supervision signal. Composite labels make this joint structure an explicit supervision target, pressuring the embedding to encode fine-grained intents that marginal losses miss.

\noindent\textbf{Inter-Task Correlation Analysis.}
We characterize task relationships via Spearman's rank correlation. Let $\mathbf{y}_i \in \mathbb{R}^N$ denote the label vector for task $i$ over $N$ samples, and let $\mathrm{rg}(\mathbf{y}_i) \in \mathbb{R}^N$ denote its rank-transformed version, the $n$-th entry of $\mathrm{rg}(\mathbf{y}_i)$ is the rank of $y_{i,n}$ among all entries of $\mathbf{y}_i$. The Spearman rank correlation between tasks $i$ and $j$ is
\begin{equation}
\rho_{ij} = \frac{\mathrm{cov}\!\bigl(\mathrm{rg}(\mathbf{y}_i),\, \mathrm{rg}(\mathbf{y}_j)\bigr)}{\sigma_{\mathrm{rg}(\mathbf{y}_i)} \cdot \sigma_{\mathrm{rg}(\mathbf{y}_j)}},
\end{equation}
where $\mathrm{cov}(\cdot,\cdot)$ denotes covariance, $\sigma_{(\cdot)}$ denotes standard deviation, and $\rho_{ij} \in [-1, 1]$ measures the strength of the monotonic relationship between the two tasks' labels. We compute the full pairwise correlation matrix $\mathbf{R} \in [-1,1]^{T \times T}$ across all $T$ tasks and apply a threshold $\tau$ to identify highly correlated task clusters $\mathcal{C}_g = \{t_{g_1}, \dots, t_{g_m}\}$, where $\rho_{t_{g_a}, t_{g_b}} \geq \tau$ for every pair within the cluster.

\noindent\textbf{MRM Label Construction.}
For each identified task group $\mathcal{C}_g$ of size $m$, we construct a composite multi-class label by Cartesian product of the individual binary labels: $\ell^{(\mathcal{C}_g)}_n = (y_{t_{g_1},n}, \dots, y_{t_{g_m},n}) \in \{0,1\}^m$, inducing a multi-class problem with up to $2^m$ classes. For example, crossing \textit{like} and \textit{comment} yields four classes $(0,0)$, $(1,0)$, $(0,1)$, and $(1,1)$, where $(1,1)$ captures a stronger expression of interest than either signal alone. For continuous-valued tasks such as time-spent, we first bucketize the numerical label into time buckets calibrated to balance the per-bucket distribution, then treat each bucket as a categorical class before applying the same Cartesian-product construction. We mine groups of size $m \in \{2, 3\}$, yielding multi-class composite tasks in the all-binary case. We use these composite tasks as supervision labels for training \mosaic{} specialists, and find MRM to be one of our most effective techniques for unique-knowledge distillation, specialists trained with MRM labels consistently surface substantial additional user signal beyond what individual task supervision provides.

\subsubsection{Cosine Redundancy Loss}

To directly enforce representational diversity, we introduce an auxiliary \emph{cosine redundancy loss} during training. Let $\mathbf{e}^{(\text{new})} \in \mathbb{R}^d$ denote the new embedding and $\mathbf{e}^{(\text{old})}_k \in \mathbb{R}^d$ the $k$-th existing  embedding. The redundancy loss is
\begin{equation}
\mathcal{L}_{\text{red}} = \frac{1}{K}\sum_{k=1}^{K} \frac{\mathbf{e}^{(\text{new})} \cdot \mathbf{e}^{(\text{old})}_k}{\|\mathbf{e}^{(\text{new})}\| \cdot \|\mathbf{e}^{(\text{old})}_k\|},
\end{equation}
and the total training objective is $\mathcal{L} = \mathcal{L}_{\text{main}} + \lambda \mathcal{L}_{\text{red}}$, where $\lambda$ controls the trade-off between task-specific accuracy and embedding orthogonality. By minimizing $\mathcal{L}_{\text{red}}$, we encourage $\mathbf{e}^{(\text{new})}$ to reside in subspaces orthogonal to the existing embeddings, maximizing the incremental information contributed to the downstream model. In practice, we tune $\lambda$ carefully, applying a warm up schedule that ramps $\lambda$ from zero so the main task converges before the redundancy penalty is enforced, this preserves orthogonality without hurting downstream NE. This formulation is a simpler analog of the redundancy reduction objectives in self-supervised representation learning~\cite{zbontar2021barlow, bardes2022vicreg}, adapted to a setting where the existing embeddings are already deployed.

\subsubsection{Additional Strategies}

We also explored two additional approaches that proved effective: (i)~\emph{feature task alignment}, where input features for the new model are selected based on feature-importance (FI) scores aligned with the specific supervision tasks, reducing feature level redundancy; and (ii)~\emph{prior embedding conditioning}, where existing  \mosaic{} embeddings are fed as input features in a training only architecture branch, enabling the new model to explicitly condition on and differentiate from prior representations.

\subsection{Optimizing Embedding Evaluation}
\label{sec:challenges:eval}

Reliable evaluation of an \mosaic{} embedding requires measuring its contribution to the downstream ranker via downstream train/eval NE. However, doing so is slow: the new embedding model must first be deployed and ramped to full  traffic, then logged into the feature pipeline to accumulate enough training data. Combined with downstream retraining and evaluation, the end-to-end iteration cycle can take several days. Upstream train/eval NE on the \mosaic{} model itself is available much earlier but does not consistently track downstream impact. In a decoupled training setup like ours, the overall upstream NE and the user-embedding quality can diverge. Only the user tower reaches downstream consumers, so improvements to the rest of the upstream graph can register as upstream NE wins without translating into downstream gains. This leaves us without a fast, accurate offline signal during day-to-day iteration and motivates the two evaluation methods described below: \emph{CoEval}, which measures the embedding's contribution through integrating the frozen user tower to the downstream ranker, and \emph{User Tower Zero-Out}, which isolates the embedding's marginal contribution within the upstream model itself.

\subsubsection{CoEval Framework}

We developed the CoEval framework as a unified architecture that measures an embedding's downstream impact through direct, in-training integration with the downstream ranker, rather than through a logged-feature proxy. The CoEval reuses the joint training architecture of the CoTrain specialist (Section~\ref{sec:system:cotrain}), but with the user tower $f_\theta$ frozen rather than trainable: for each downstream training sample $(x_u, x_c, y)$, the CoEval computes embedding $\mathbf{e}_u = f_\theta(x_u)$ inline and feeds it into the downstream model alongside the other features. The resulting NE delta directly quantifies the embedding's contribution,
\begin{equation}
\Delta \text{NE}_{\text{CoEval}} = \text{NE}_{\text{with}\;\mathbf{e}_u} - \text{NE}_{\text{baseline}}.
\end{equation}
The design has two key properties. First, the CoEval produces a downstream-aligned eval signal directly from a trained specialist checkpoint, without requiring deployment of the specialist (config rollout, model loading, traffic ramp) nor logging and accumulation of its embeddings in the downstream feature pipeline. A freshly trained user tower can be evaluated immediately, even before the specialist is wired into the feature stack. Second, the CoEval is modular: it treats the user tower as a frozen module that pairs with a copy of any downstream ranker, so any specialist can be evaluated this way, not only CoTrain specialists. As a practical consequence, the end-to-end evaluation cycle compresses down to less than half of its original time. Validation experiments against the full feature logging pipeline show that the CoEval produces accurate evaluations aligned with downstream NE.

\subsubsection{User Tower Zero-Out Evaluation}

We also developed User Tower Zero-Out as an alternative method for accurately evaluating user-embedding quality without the feature-logging step. This technique measures the $\Delta$NE produced by zeroing out the user-tower output in the \mosaic{} model, providing a gauge on the incremental value of the embedding signal within the upstream context.

% =============================================================================
% Section 5: Experiments
% Setup / Offline / SID / Online / Ablation / Deployment-narrative.
% =============================================================================

\section{Experiments}
\label{sec:deployment}

\subsection{Experimental Setup}
\label{sec:deployment:setup}

\noindent\textbf{Datasets and tasks.}
We evaluated \mosaic{} on multiple downstream  rankers on various surfaces, as listed in Table~\ref{tab:mosaic-ne-wins}. Each downstream ranker optimizes its own set of tasks and labels, and we report task-level $\Delta$NE measured on the same train/eval split used by the  ranker, where task A represents the top priority task while task B/C are other engagement events.

\noindent\textbf{Offline Metrics.}
We use \emph{NE} and \emph{FI} as our gold-standard offline evaluation metrics. NE is the log-loss-based quality metric used by our downstream rankers; lower is better. FI measures the relative ranking of the feature \mosaic{} among all input features in the downstream ranker.

\noindent\textbf{Online Testing.}
We also conducted online A/B tests to measure the contribution of \mosaic{} embeddings on user engagement and topline metrics in different surfaces.

\subsection{Offline Evaluation}
\label{sec:deployment:offline}

\noindent\textbf{Offline eval.}
Table~\ref{tab:mosaic-ne-wins} summarizes the downstream offline NE gains on selected tasks. We observe consistent and statistically significant gains from \mosaic{} across multiple surfaces.

\begin{table}[h]
  \caption{\mosaic{} embedding $\Delta$NE on downstream  surfaces. Lower (more negative) is better.}
  \label{tab:mosaic-ne-wins}
  \centering
  \small
  \begin{tabular}{@{}llrr@{}}
  \toprule
  \textbf{Surface Name} & \textbf{Multi-Task Head} & \textbf{Train $\Delta$NE} & \textbf{Eval $\Delta$NE} \\
  \midrule
  \multirow{3}{*}{Surface 1}
    & Task A       & $-$0.37\%  & $-$0.21\%  \\
    & Task B          & $-$0.38\%  & $-$0.30\%  \\
    & Task C         & $-$0.30\%  & $-$0.16\%  \\
  \midrule
  \multirow{3}{*}{Surface 2}
    & Task A       & $-$0.03\%  & $-$0.03\%  \\
    & Task B          & $-$0.04\%  & $-$0.05\%  \\
    & Task C       & $-$0.07\%  & $-$0.02\%  \\
  \midrule
  \multirow{2}{*}{Surface 3}
    & Task A         & $-$0.10\%  & $-$0.05\%  \\
    & Task B       & $-$0.05\%  & $-$0.21\%  \\
  \midrule
  \multirow{3}{*}{Surface 4}
    & Task A              & $-$0.27\%  & $-$0.13\%  \\
    & Task B    & $-$0.31\%  & $-$0.21\%  \\
    & Task C & $-$0.31\%  & $-$0.16\%  \\
  \midrule
  \multirow{2}{*}{Surface 5}
    & Task A            & $-$0.09\%  & $-$0.11\%  \\
    & Task B       & $-$0.03\%  & $-$0.08\%  \\
  \midrule
  \multirow{2}{*}{Surface 6}
    & Task A     & $-$0.05\%  & $-$0.09\%  \\
    & Task B       & $-$0.21\%  & $-$0.33\%  \\
  \bottomrule
  \end{tabular}
\end{table}

\noindent\textbf{Specialist family ablation.}
We also conducted an ablation study to break down the NE contribution across specialist families. Table~\ref{tab:specialist_contribution} illustrates the breakdown on Surface 1: dense heavy specialists, with $\sim$20 embeddings deployed, contribute the largest share at $-$0.22\% NE, followed by CoTrain ($-$0.15\%) and HSTU-CInt sequential ($-$0.12\%) models.

\begin{table}[t]
\caption{NE contribution breakdown across \mosaic{} specialist families.}
\label{tab:specialist_contribution}
\small
\centering
\begin{tabular}{@{}lcc@{}}
\toprule
\textbf{Specialist Family} & \textbf{\#~Added Embeddings} & \textbf{Aggregated Eval $\Delta$NE} \\
\midrule
Memorization-driven   & 8  & -0.09\%      \\
Dense-heavy           & 20  & -0.22\%      \\
Sequential-based      &  6  & -0.12\% \\
CoTrain              &  3  & -0.15\% \\
\bottomrule
\end{tabular}
\end{table}

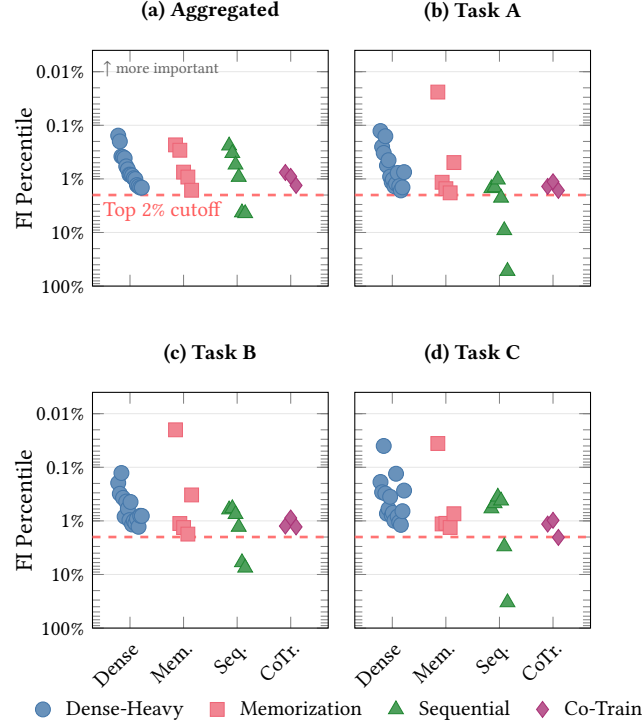
\begin{figure}[t]
\centering
\begin{tikzpicture}
\begin{groupplot}[
  group style={
    group size=2 by 2,
    horizontal sep=10pt,
    vertical sep=40pt,
    ylabels at=edge left,
    yticklabels at=edge left,
    xlabels at=edge bottom,
    xticklabels at=edge bottom,
  },
  width=4.7cm,
  height=4.7cm,
  ymode=log,
  y dir=reverse,
  ymin=1, ymax=25000,
  xmin=0.3, xmax=4.7,
  xtick={1,2,3,4},
  xticklabels={Dense, Mem., Seq., CoTr.},
  x tick label style={font=\small, rotate=45, anchor=north east},
  ytick={2.5, 25, 250, 2500, 25000},
  yticklabels={0.01\%, 0.1\%, 1\%, 10\%, 100\%},
  ymajorgrids=true,
  xmajorgrids=false,
  grid style={gray!20, line width=0.3pt},
  tick label style={font=\small},
  ylabel={FI Percentile},
  ylabel style={font=\normalsize, at={(axis description cs:-0.22,0.5)}},
  title style={font=\small\bfseries, at={(0.5,1.02)}, anchor=south},
  every axis plot/.append style={mark options={solid}},
  clip=true,
]

%%% (a) Aggregated %%%
\nextgroupplot[
  title={(a) Aggregated},
  legend to name=sharedlegend,
  legend style={
    font=\small, draw=none, legend columns=4,
    column sep=4pt, row sep=2pt,
    /tikz/every even column/.append style={column sep=8pt},
  },
]
% Top 2% cutoff line
\draw[dashed, red!55, line width=1.0pt] (axis cs:0.3,500) -- (axis cs:4.7,500);
\node[font=\small, red!65, anchor=south east, inner sep=2pt]
  at (axis cs:2.8,2000) {Top 2\% cutoff};
  
% "higher = more important" annotation
\node[font=\scriptsize, black!60, anchor=north west, inner sep=2pt]
  at (axis description cs:0.02,0.98) {$\uparrow$ more important};
% Dense-Heavy
\addplot[only marks, mark=*, cdense, fill=cdense!80, mark size=2.8pt] coordinates {
  (0.78,39)(0.81,50)(0.84,95)(0.87,98)(0.90,103)(0.93,146)(0.96,166)(0.99,208)
  (1.01,211)(1.04,217)(1.07,243)(1.10,252)(1.13,326)(1.16,344)(1.19,363)(1.22,364)
};
\addlegendentry{Dense-Heavy}
% Memorization
\addplot[only marks, mark=square*, cmem, fill=cmem!80, mark size=2.6pt] coordinates {
  (1.85,58)(1.93,73)(2.00,187)(2.08,231)(2.15,407)
};
\addlegendentry{Memorization}
% Sequential
\addplot[only marks, mark=triangle*, cseq, fill=cseq!80, mark size=3.2pt] coordinates {
  (2.85,60)(2.91,83)(2.97,136)(3.03,231)(3.09,1061)(3.15,1098)
};
\addlegendentry{Sequential}
% Co-Train
\addplot[only marks, mark=diamond*, cco, fill=cco!80, mark size=3.0pt] coordinates {
  (3.90,190)(4.00,225)(4.10,331)
};
\addlegendentry{Co-Train}

%%% (b) Comment %%%
\nextgroupplot[title={(b) Task A}]
\draw[dashed, red!55, line width=1.0pt] (axis cs:0.3,500) -- (axis cs:4.7,500);
\addplot[only marks, mark=*, cdense, fill=cdense!80, mark size=2.8pt, forget plot] coordinates {
  (0.78,32)(0.81,63)(0.84,82)(0.87,40)(0.90,141)(0.93,112)(0.96,227)(0.99,288)
  (1.01,256)(1.04,343)(1.07,307)(1.10,194)(1.13,341)(1.16,410)(1.19,360)(1.22,187)
};
\addplot[only marks, mark=square*, cmem, fill=cmem!80, mark size=2.6pt, forget plot] coordinates {
  (1.85,6)(1.93,289)(2.00,381)(2.08,455)(2.15,124)
};
\addplot[only marks, mark=triangle*, cseq, fill=cseq!80, mark size=3.2pt, forget plot] coordinates {
  (2.85,375)(2.91,360)(2.97,259)(3.03,565)(3.09,2279)(3.15,13136)
};
\addplot[only marks, mark=diamond*, cco, fill=cco!80, mark size=3.0pt, forget plot] coordinates {
  (3.90,346)(4.00,283)(4.10,410)
};

%%% (c) Like %%%
\nextgroupplot[title={(c) Task B}]
\draw[dashed, red!55, line width=1.0pt] (axis cs:0.3,500) -- (axis cs:4.7,500);
\addplot[only marks, mark=*, cdense, fill=cdense!80, mark size=2.8pt, forget plot] coordinates {
  (0.78,49)(0.81,78)(0.84,32)(0.87,93)(0.90,208)(0.93,109)(0.96,148)(0.99,242)
  (1.01,111)(1.04,292)(1.07,251)(1.10,278)(1.13,229)(1.16,322)(1.19,201)(1.22,202)
};
\addplot[only marks, mark=square*, cmem, fill=cmem!80, mark size=2.6pt, forget plot] coordinates {
  (1.85,5)(1.93,280)(2.00,333)(2.08,441)(2.15,82)
};
\addplot[only marks, mark=triangle*, cseq, fill=cseq!80, mark size=3.2pt, forget plot] coordinates {
  (2.85,154)(2.91,146)(2.97,188)(3.03,325)(3.09,1486)(3.15,1876)
};
\addplot[only marks, mark=diamond*, cco, fill=cco!80, mark size=3.0pt, forget plot] coordinates {
  (3.90,312)(4.00,221)(4.10,324)
};

%%% (d) Share %%%
\nextgroupplot[title={(d) Task C}]
\draw[dashed, red!55, line width=1.0pt] (axis cs:0.3,500) -- (axis cs:4.7,500);
\addplot[only marks, mark=*, cdense, fill=cdense!80, mark size=2.8pt, forget plot] coordinates {
  (0.78,47)(0.81,73)(0.84,10)(0.87,77)(0.90,183)(0.93,149)(0.96,89)(0.99,198)
  (1.01,177)(1.04,249)(1.07,33)(1.10,207)(1.13,265)(1.16,297)(1.19,165)(1.22,68)
};
\addplot[only marks, mark=square*, cmem, fill=cmem!80, mark size=2.6pt, forget plot] coordinates {
  (1.85,9)(1.93,285)(2.00,278)(2.08,332)(2.15,184)
};
\addplot[only marks, mark=triangle*, cseq, fill=cseq!80, mark size=3.2pt, forget plot] coordinates {
  (2.85,147)(2.91,118)(2.97,89)(3.03,105)(3.09,751)(3.15,8313)
};
\addplot[only marks, mark=diamond*, cco, fill=cco!80, mark size=3.0pt, forget plot] coordinates {
  (3.90,287)(4.00,244)(4.10,509)
};

\end{groupplot}

% Shared legend below
\path (group c1r2.south) -- (group c2r2.south) coordinate[midway] (legendmid);
\node[below=20pt] at (legendmid) {\ref{sharedlegend}};

\end{tikzpicture}
\caption{The FI breakdown of \mosaic{} among specialist categories.}
\Description{Four scatter panels in a two-by-two grid reporting feature importance (FI) percentile for individual \mosaic{} embeddings, grouped by specialist category. Panel (a) shows the aggregated ranking across all downstream ranking tasks; panels (b), (c), and (d) show three individual tasks labeled Task A, Task B, and Task C. In every panel the horizontal axis lists the four specialist categories---Dense-Heavy, Memorization, Sequential, and Co-Train---and the vertical axis is FI percentile on a reversed logarithmic scale running from 0.01\% at the top to 100\% at the bottom, so that higher points denote more important features. A red dashed horizontal line marks the top 2\% cutoff. In the aggregated panel, every specialist category places all or nearly all of its embeddings above this cutoff: dense-heavy specialists form a tight, consistent band; a handful of memorization-driven embeddings reach the very top of the chart near the 0.01\% percentile; sequential specialists sit high but spread more widely, with two embeddings falling just below the cutoff; and Co-Train embeddings cluster within the upper band. The per-task panels show the same overall pattern with greater dispersion, the sequential category exhibiting the widest spread and a few embeddings ranking well below the cutoff on individual tasks.}
\label{fig:feature_importance}
\end{figure}

\noindent\textbf{Feature importance.}
We also evaluated individual \mosaic{} embeddings through FI on downstream ranking models. Figure~\ref{fig:feature_importance} illustrates the FI of various \mosaic{} embeddings on a production surface, showing the ranking for three of the most important tasks and an \emph{aggregated} ranking that considers all 10+ ranking tasks used in the downstream ranker. Among the 20k+  input features, \mosaic{} embeddings consistently rank in the top 2\%. A few memorization-driven embeddings rank near the very top ($\sim$0.01\%), while dense-heavy specialists deliver on average very high and consistent FI performance.

% Feature importance

\noindent\textbf{Sequential scaling ablation.}
\label{sec:deployment:seq_scaling}
We reports NE and QPS deltas across three scaling axes for the sequential specialist (Section~\ref{sec:system:fleet}). Using a 512 sequence length as the baseline, we experiment with varying sequence length, sparse MoE routing, and embedding dimensions. Table~\ref{tab:seq_scaling} highlights the performance to capacity tradeoffs in scaling across each category.

\begin{table}[t]
\caption{Sequential model scaling: NE improvement on tasks across Surfaces 1 and 2.}
\label{tab:seq_scaling}
\small
\centering
\begin{tabular}{@{}llcc@{}}
\toprule
\textbf{Category} & \textbf{Configuration} & \textbf{Eval $\Delta$NE} & \textbf{QPS} \\
\midrule
\multirow{2}{*}{Seq length}
 & HSTU, 1024 seq len & $-$0.21\% & $-$18\% \\
 & HSTU, 2048 seq len & $-$0.45\% & $-$33\% \\
\midrule
\multirow{4}{*}{MoE}
 & top-2 of 8 experts, expert\_dim=128 & $-$0.07\% & $-$8\% \\
 & top-2 of 8 experts, expert\_dim=256 & $-$0.12\% & $-$12\% \\
 & top-2 of 8 experts, expert\_dim=512 & $-$0.14\% & $-$21\% \\
 & top-4 of 8 experts, expert\_dim=512 & $-$0.15\% & $-$23\% \\
\midrule
\multirow{2}{*}{Emb dim scale-up}
 & emb\_size=256 & $-$0.23\% & $-$15\% \\
 & emb\_size=256 + warmup & $-$0.25\% & $-$18\% \\
\bottomrule
\end{tabular}
\end{table}

\subsection{Online A/B Results}
\label{sec:deployment:online}

\begin{table}[t]
  \caption{Topline online A/B lifts from \mosaic{} embeddings on downstream  surfaces.}
  \label{tab:online_ab}
  \centering
  \small
  \begin{tabular}{@{}lr@{}}
  \toprule
  \textbf{Product Surface} & \textbf{Topline Gains $\Delta$} \\
  \midrule
  Surface 1 & $+$0.10\% \\
  Surface 2        & $+$0.15\% \\
  Surface 3     & $+$0.28\% \\
  \bottomrule
  \end{tabular}
\end{table}

We validated downstream impact through online A/B experiments across multiple  surfaces, with a few topline lifts reported in Table~\ref{tab:online_ab}. \mosaic{} delivers consistent and statistically significant lifts on every surface tested.

% =============================================================================
% Section 8: Conclusion
% =============================================================================

\section{Conclusion}
\label{sec:conclusion}

We presented \mosaic{}, a foundational user-modeling platform that organizes user representation as a managed \emph{fleet of specialist user embedding models} --- memorization-driven, dense-heavy, sequential, and CoTrain --- each architecturally matched to a distinct facet of user behavior. To keep the library sustainable as it grows, we introduced \emph{Multi-Task Relation Mining} (MRM) and a \emph{cosine redundancy loss} to maximize unique user-knowledge distillation across the fleet, a logging-free embedding evaluation pipeline (CoEval and User Tower Zero-Out) that improves iteration velocity by 3--5$\times$, and a hybrid CPU/GPU, online-and-offline serving stack. \mosaic{} delivers consistent offline NE improvements and statistically significant online A/B engagement lifts. The design principles and operating mechanisms behind \mosaic{} provides a practical blueprint for industrial recommendation systems looking to scale user representations across many downstream consumers.

% =============================================================================
% Acknowledgments — replace placeholder with real text before submission.
% =============================================================================
\begin{acks}
We would like to thank all the contributions from the \mosaic{} team and partners who collaborated with us to expand the impact of \mosaic{}: Ryan Shue, Han Jiang, Wanqiang Chen, Yang Liu, Hanxiong Chen, Zhuoran Yu, Haoran Ye, Lynette Gao, Li Lu, Charles Hubbard, Yan Sun, Brian Choi, Deepa Rao, Zitong Zeng, Cheng Huang, Andy Wang, Zhao Zhu, Pengchao Wang, Chenyu Zhang, Mingwei Hu, Lu Zhang, Han Li, Daisy Shi He, Xin Zhuang, Guoqiang Jerry Chen, Han Lu, Xuesong Hou, Giri Rajaram, Hsiao-Ping Tseng, Mark Gluzman, Leo Liu, Shishuang He, Wei Sun, Thanh Dang, Chao Wang, Eric Landgrebe, Farzaneh Rajabi, Grace (Guangchao) Yuan, Henry A Lin, Jiayan Gan, Jieqian He, Sifan Liu, Lebo Wang, Shalmoli Gupta.
\end{acks}

% =============================================================================
% Author Biographies (RecSys 2026 Industry Track expects short bios per author).
% Add a one-paragraph bio per author here when the author list is final.
% =============================================================================

\bibliographystyle{ACM-Reference-Format}
\bibliography{references}

\end{document}